\documentclass[a4paper,fleqn]{cas-dc}
\usepackage[version=4]{mhchem}
\usepackage{amsmath}
\usepackage{hyperref}
\usepackage[authoryear,longnamesfirst]{natbib}
\usepackage{lineno}
\usepackage{adjustbox}
\def\tsc#1{\csdef{#1}{\textsc{\lowercase{#1}}\xspace}}
\tsc{WGM}
\tsc{QE}
\tsc{EP}
\tsc{PMS}
\tsc{BEC}
\tsc{DE}

\begin{document}
	\let\WriteBookmarks\relax
	\def\floatpagepagefraction{1}
	\def\textpagefraction{.001}
	\shorttitle{Amino acetonitrile towards G10.47+0.03}
	\shortauthors{Manna \& Pal}
	\title [mode = title]{Identification of interstellar amino acetonitrile in the hot molecular core G10.47+0.03: Possible glycine survey candidate for the future}  
	
	\author[1]{Arijit Manna}
	\address[1]{Midnapore City College, Kuturia, Bhadutala, Paschim Medinipur, West Bengal, India 721129}
	
	\author[1]{Sabyasachi Pal}
	\cormark[1]

	\begin{abstract}
	Amino acids are the essential keys that contribute to the study of the formation of life. The simplest amino acid, glycine (\ce{NH2CH2COOH}), has been searched for a long time in the interstellar medium, but all surveys of glycine have failed. Since the detection of glycine in the interstellar medium was extremely difficult, we aimed to search for the precursor of glycine. After detailed searches of the individual prebiotic molecular species, we successfully identified the emission lines of possible glycine precursor molecule amino acetonitrile (\ce{NH2CH2CN}) towards the hot molecular core G10.47+0.03 using the Atacama Large Millimeter/Submillimeter Array. We estimated the statistical column density of amino acetonitrile was (9.10$\pm$0.7)$\times$10$^{15}$ cm$^{-2}$ with rotational temperature ($T_{rot}$) 122$\pm$8.8 K. The estimated fractional abundance of amino acetonitrile was 7.01$\times$10$^{-8}$. We found that the estimated fractional abundance of \ce{NH2CH2CN} fairly agrees with the theoretical value predicted by the three-phase warm-up model from \cite{gar13}.
		
	\end{abstract}
	 
	\maketitle
	
\section{Introduction}
\label{sec:intro} 	
 In the interstellar medium, more than 260 molecules have been detected using the millimeter/submillimeter radio telescopes\footnote{\href{https://cdms.astro.uni-koeln.de/classic/molecules}{https://cdms.astro.uni-koeln.de/classic/molecules}}. The hot molecular cores are chemically rich complex systems in the interstellar medium, and several complex organic molecules have been detected in the hot molecular core regions \citep{gar17,bel16,gor20,man21}. In this article, we give priority to the study of prebiotic organic molecular spectral lines towards the hot molecular core region G10.47+0.03 (hereafter, G10), which was observed at a distance of 8.6 kpc \citep{san14}. The luminosity of the hot molecular core G10 was 10$^{6}$ L$_{\odot}$, which indicates that G10 is one of the highest luminosity star-forming region in the galaxy and particularly interesting to the investigation of the molecular lines \citep{rol09}. The emission lines of \ce{NH2CHO}, \ce{CH3NCO}, and HNCO were detected towards G10 using the Atacama Large Millimeter/Submillimeter Array \citep{gor20}. Recently, the complex molecular emission lines of CH$_{3}$OH, (CH$_{2}$OH)$_{2}$, CH$_{3}$CHO, CH$_{3}$CH$_{2}$CHO, HOCH$_{2}$CHO, CH$_{3}$COCH$_{3}$, CH$_{3}$OCHO, and CH$_{3}$OCH$_{3}$ were also detected towards G10 using the ALMA \citep{mondal21}. Earlier, the emission lines of prebiotic complex molecules methanimine (\ce{CH2NH}) and methylamine (\ce{CH3NH2}) were detected towards G10, which are the simplest imine and amine, respectively \citep{suz16, ohi19}. The prebiotic molecule \ce{CH3NH2} plays an important role in the formation of the glycine isomer \ce{CH3NHCOOH} towards the hot molecular cores \citep{hol05}. The bio-molecule \ce{CH3NH2} will be produced during the warm-up phases between the reactions of \ce{NH2} and \ce{CH3} under cosmic ray irradiation on the grain surface (\ce{NH2}+\ce{CH3}$\longrightarrow$\ce{CH3NH2}) \citep{kim11}. When \ce{CH3NH2} molecule reacts with \ce{CO2} under UV irradiation, it forms the glycine isomer molecule \ce{CH3NHCOOH} in the solid phase (\ce{CH3NH2}$\stackrel{\rm CO_{2}}{\longrightarrow}$\ce{CH3NHCOOH}) \citep{hol05}. In another way, \ce{CH3NH2} molecule can be produced in the interstellar medium due to hydrogenation of HCN on the dust surface of hot molecular cores (HCN $\stackrel{\rm 2H}{\longrightarrow}$ CH$_2$NH$\stackrel{\rm 2H}{\longrightarrow}$CH$_3$NH$_2$) \citep{thu11}. As two possible precursors of glycine \ce{CH3NH2} and \ce{CH2NH} were already detected towards G10 \citep{ohi19, suz16}, we tried to search for another prebiotic molecule, amino acetonitrile (\ce{NH2CH2CN}) towards G10, which was known as another possible precursor of glycine \citep{bro77, eli07, ohi19}.

\begin{figure*}
	\includegraphics[width=0.98\textwidth]{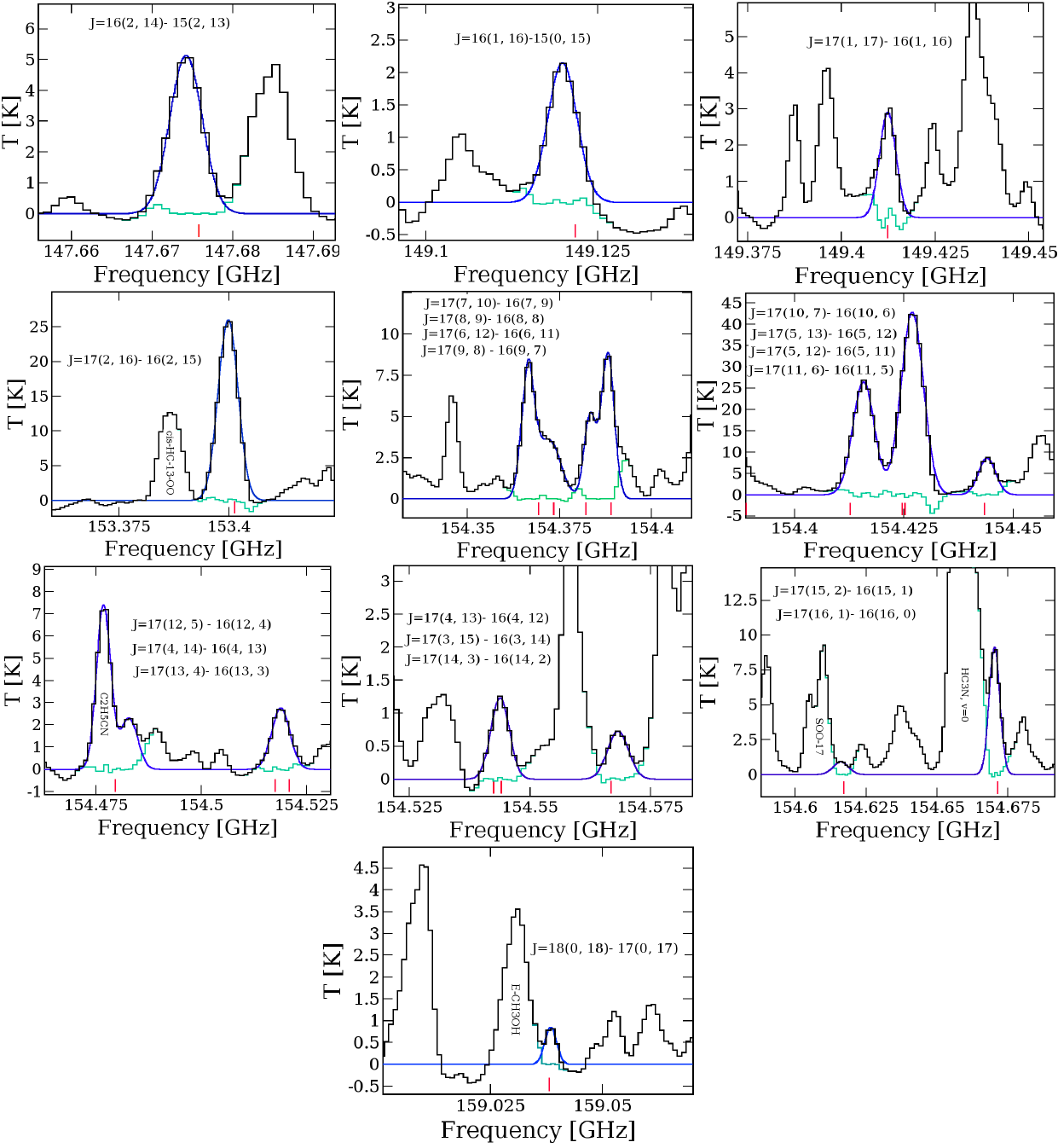}
	\caption{Rotational emission spectra of AAN with different molecular transitions towards G10. From the emission spectra, the continuum emission was completely subtracted. The black line showed the observed spectra, while the blue line showed the Gaussian model, which was overlaid on the observed spectra of AAN. The green spectra indicated the residual. The red lines indicate the peak position of the detected transitions of AAN.}
	\label{fig:emission} 
\end{figure*}

 The complex nitrile molecule amino acetonitrile (hereafter, AAN) is known as one of the important rare compounds in the interstellar medium, which is also known as glycine nitrile \citep{wis07}. The AAN molecule is one of the important species for astrochemists and astrobiologists because AAN can be converted into glycine after hydrolysis via glycinamide (\ce{NH2CH2(O)NH2}) \citep{alo18, pel84, wis07}. Earlier, astrochemists suggested the possible chemical pathways for the formation of glycine from AAN via glycinamide on the dust surface of the hot molecular cores:\\\\
HCN$\stackrel{\rm 2H}{\longrightarrow}$CH$_{2}$NH~~~\citep{thu11, woon02}~~~~~(1)\\
CH$_{2}$NH$\stackrel{\rm HCN}{\longrightarrow}$NH$_{2}$CH$_{2}$CN \citep{dan11}~~~~~~~~~~~~~~~~~~~~~~~~~~~~~~~(2)\\ 
NH$_{2}$CH$_{2}$CN$\stackrel{\rm H_{2}O}{\longrightarrow}$ \ce{NH2CH2C(O)NH2} \citep{alo18}~~~~~~(3)\\
\ce{NH2CH2C(O)NH2}$\stackrel{\rm H_{2}O}{\longrightarrow}$NH$_{2}$CH$_{2}$COOH \citep{alo18}(4)\\\\
 Reaction 1 indicated that the \ce{CH2NH} molecule was produced on the dust surface via the hydrogenation of HCN \citep{thu11, woon02}. Reaction 2 presented the formation pathways of AAN between the reactions of CH$_{2}$NH and HCN via the Strecker synthesis reaction \citep{dan11}. Now, in the interstellar medium, the hydrolysis of AAN in the gas phase or the icy mantle on the grain surface produces glycinamide (\ce{NH2CH2C(O)NH2}), as shown in reaction 3 \citep{alo18}. The hydrolysis of glycinamide forms glycine on the grain surface, which was presented in reactions 4 \citep{alo18}. Recently, \citet{kis22} searched for the emission lines of glycinamide from Sgr B2(N) but they cannot find any evidence of glycinamide within the limit of the Local Thermodynamic Equilibrium (LTE) model. They estimated the upper limit column density of glycinamide towards Sgr B2(N) was $\leqslant$1.5$\times$10$^{16}$ cm$^{-2}$. Earlier, \citet{koc08} found that \ce{H2O} can efficiently catalyze a reaction between \ce{CH2NH} and HNC to form AAN in the grain mantles at a temperature of 50 K. The AAN molecule contains both amino and nitrile groups \citep{sharma}. The survival of AAN against UV photolysis in the interstellar ices was $\sim$5 times longer than glycine \citep{ber04}. Earlier, \citet{wis07} did not found the molecular lines of AAN from hot molecular cores Orion KL, W51 e1/e2, S140, and W3(OH) using the Onsala 20 m telescope. The emission lines of interstellar AAN were first time detected towards Sgr B2(N) with column density 2.8$\times$10$^{16}$ cm$^{-2}$ and excitation temperature 100 K using IRAM, and later the emission lines of AAN in Sgr B2(N) was verified by ATCA and VLA \citep{bel08}. Recently, astrochemists tried to search for the emission lines of AAN and glycine towards the quiescent giant molecular cloud G+0.693--0.027 and hot corino IRAS 16293--2422 B but they did not detect them within the LTE approximation \citep{jim20}. After the non detection of AAN from G+0.693--0.027 and IRAS 16293--2422 B, \citet{jim20} estimated the upper limit column density of AAN in G+0.693--0.027 and IRAS 16293--2422 B was $\leq$0.6$\times$10$^{13}$ cm$^{-2}$ and $\leq$1.2$\times$10$^{14}$ cm$^{-2}$ respectively. The electric dipole moment of AAN is $\mu$$_{a}$ = 2.577 D and $\mu$$_{b}$ = 0.5754 D which indicates that the a-type transitions are 20 times higher than b-types transitions \citep{pic, sharma}.

\begin{table*}
	\centering
	\caption{Summary of the line parameters of the AAN towards G10.}
	\begin{adjustbox}{width=1.0\textwidth}
		\begin{tabular}{ccccccccccccccccc}
			\hline 
			Frequency range&Observed frequency &Transition & $E_{u}$ & $A_{ij}$ &Peak intensity&S$\mu^{2}$&FWHM &$\rm{\int T_{mb}dV}$ & V$_{LSR}$ &Remark\\
			
			(GHz)&(GHz) &(${\rm J^{'}_{K_a^{'}K_c^{'}}}$--${\rm J^{''}_{K_a^{''}K_c^{''}}}$) &(K)&(s$^{-1}$) &(K)&(Debye$^{2}$)& (km s$^{-1}$) &(K km s$^{-1}$)&(km s$^{-1}$) & \\
			\hline
			147.55-148.01$^{\odot}$&147.6758&16(2,14)--15(2,13)$^{\dagger}$&64.76&1.19$\times$10$^{-4}$&5.081&104.623&5.262$\pm$0.62&40.629$\pm$4.56&68.607&blended with c-HCC$^{13}$CH\\
			\hline
			148.50--149.43$^{\circledast}$&149.1218&16(1,16)--15(0,15)&58.79&4.50$\times$10$^{-6}$&2.123&3.8437&6.653$\pm$0.98&20.824$\pm$3.16&68.820&blended with c-\ce{C3H2}\\
			
			&149.4123&17(1,17)--16(1,16)&65.96&1.25$\times$10$^{-4}$&3.014&112.419&4.582$\pm$0.32&31.425$\pm$2.56&68.506&\bf{Detected}\\
			\hline
			153.00--153.93$^{\circledast}$&153.4012&17(2,16)--16(2,15)&71.37&1.34$\times$10$^{-4}$&25.778&111.262&6.286$\pm$0.97&232.587$\pm$12.69&68.589&blended with \ce{CH3CHO}\\
			\hline
			154.0--154.93$^{\circledast}$&154.3692&17(7,10)--16(7,9)$^{\dagger*}$&127.05&1.15$\times$10$^{-4}$&4.753&93.746&6.975$\pm$0.63&72.498$\pm$10.32&68.692&blended with c-HCC$^{13}$CN\\
			
			&154.3734&17(8,9)--16(8,8)$^{\dagger*}$&145.50&1.08$\times$10$^{-4}$&3.234&87.892&5.628$\pm$0.78&96.558$\pm$19.39&68.596&blended with \ce{SO2}\\
			
			&154.3822&17(6,12)--16(6,11)$^{\dagger*}$&111.04&1.021$\times$10$^{-4}$&4.645&98.835&5.698$\pm$0.43&34.261$\pm$3.97&68.521&blended with SO$^{18}$O and \ce{NH2D}\\
			
			&154.3889&17(9,8)--16(9,7)$^{\dagger*}$&146.41&9.95$\times$10$^{-5}$&8.984&81.250&5.889$\pm$0.97&103.936$\pm$20.39&68.561&blended with E-\ce{CH3OH}\\
			
			&154.4128&17(10,7)--16(10,6)$^{\dagger*}$&189.75&9.04$\times$10$^{-5}$&26.713&73.837&6.823$\pm$1.26&258.351$\pm$42.93&68.643&blended with HNCO\\
			
			&154.4246&17(5,13)--16(5,12)$^{\dagger}$&97.50&1.26$\times$10$^{-4}$&42.224&103.125&6.956$\pm$0.86&321.989$\pm$60.29&68.597&blended with  E-\ce{CH3OH}\\
			
			&154.4252&17(5,12)--16(5,11)$^{\dagger}$&97.50&1.26$\times$10$^{-4}$&42.224&103.124&6.958$\pm$0.87&321.998$\pm$60.35&68.597&blended with  E-\ce{CH3OH}\\
			
			&154.4433&17(11,6)--16(11,5)$^{\dagger*}$&215.52&8.04$\times$10$^{-5}$&8.783&65.630&4.805$\pm$0.69&64.882$\pm$10.36&68.563&blended with \ce{HC3N}\\
			
			&154.4796&17(12,5)--16(12,4)$^{\dagger*}$&243.71&6.95$\times$10$^{-5}$&2.814&56.638&4.207$\pm$0.67&46.244$\pm$12.16&68.543& blended with \ce{C2H5CN}\\
			
			&154.5175&17(4,14)--16(4,13)$^{\dagger}$&86.42&1.31$\times$10$^{-4}$&2.695&106.637&4.221$\pm$0.34&33.667$\pm$2.96&68.537&\bf{Detected}\\
			&154.5209&17(13,4)--16(13,3)$^{\dagger*}$&274.31&5.75$\times$10$^{-5}$&2.695&46.879&4.221$\pm$0.34&33.667$\pm$2.94&68.537&\bf{Detected}\\
			
			&154.5424&17(4,13)--16(4,12)$^{\dagger}$&86.43&1.31$\times$10$^{-4}$&1.269&106.653&4.821$\pm$0.98&66.107$\pm$5.16&68.594&blended with U line\\
			
			&154.5440&17(3,15)--16(3,14)$^{\dagger}$&77.81&1.34$\times$10$^{-4}$&1.269&109.368&4.821$\pm$0.98&66.107$\pm$5.19&68.594& blended with U line\\
			
			&154.5666&17(14,3)--16(14,2)$^{\dagger*}$&307.31&4.46$\times$10$^{-5}$&0.758&36.333&4.536$\pm$0.78&97.258$\pm$9.83&68.439&\bf{Detected}\\
			
			&154.6170&17(15,2)--16(15,1)$^{\dagger*}$&342.71&3.07$\times$10$^{-5}$&0.953&25.001&4.789$\pm$0.62&8.595$\pm$2.69&68.396&blended with c-\ce{C3H}\\
			
			&154.6714&17(16,1)--16(16,0)&380.48&1.59$\times$10$^{-5}$&8.863&12.892&5.537$\pm$0.68&86.402$\pm$12.98&68.513&blended with \ce{HC3N}\\
			\hline
			158.49--159.43$^{\circledast}$&159.0381&18(0,18)--17(0,17)&73.29&1.51$\times$10$^{-4}$&0.814&119.204&4.823$\pm$0.32&3.791$\pm$0.97&68.509&\bf{Detected}\\
			
			\hline
		\end{tabular}	
	\end{adjustbox}
	\\
	${{\dagger}}$ -- Those transitions of AAN were also detected in Sgr B2 \citep{bel08}.\\
	${*}$ -- Transitions contain double with frequency difference less than 100 kHz. The quantum numbers of the second are not shown.\\
	${\odot}$-- Spectral resolution 488.28 kHz\\
	$\circledast$ -- Spectral resolution 1128.91 kHz\\
	
	\label{tab:MOLECULAR DATA}
\end{table*}

In this article, we presented the first interferometric detection of complex amino and nitrile mixed molecule AAN towards the hot molecular core G10 using ALMA. We used the rotational diagram method under the LTE conditions to calculate the column density ($N$) and rotational temperature ($T_{rot}$) of the emission lines of AAN. We also compared our estimated abundance of AAN with the three-phase warm-up model of \citet{gar13}. The ALMA observation and data reduction were presented in Section~\ref{obs}. The result of the detection of emission lines of AAN was shown in Section~\ref{ref}. The discussion and conclusion of the detection of AAN  were shown in Section~\ref{dis} and \ref{conclu}. 

\section{Observations and data reductions}
\label{obs}
 The hot molecular core G10 was observed to study the interstellar complex organic prebiotic molecules using the high-resolution Atacama Large Millimeter/submillimeter Array (ALMA)\footnote{\href{https://almascience.nao.ac.jp/asax/}{https://almascience.nao.ac.jp/asax/}} with band 4 in cycle 4. The hot molecular core G10 was observed on 28th January 2017, 5th March 2017, 6th March 2017, and 7th March 2017 with phase center of ($\alpha,\delta$)$_{\rm J2000}$ = (18:08:38.232, --19:51:50.400). The on-source integration time on these days was 2026.080 sec, 3810.240 sec, 6804.000 sec, and 1723.680 sec, respectively, and a total of thirty-nine, forty, forty-one, and thirty-nine antennas were used during these observations. The observation aimed to study five-strong emission lines of interstellar glycine towards G10. The observations of G10 were carried out in four spectral windows with frequency ranges 129.50--134.44 GHz, 147.50--149.43 GHz, 153.00--154.93 GHz, and 158.49--160.43 GHz with a spectral resolution of 1128.91 kHz and 488.28 kHz. The angular resolution of different frequency ranges was 1.67$^{\prime\prime}$ (14362 AU), 1.52$^{\prime\prime}$ (13072 AU), 1.66$^{\prime\prime}$ (14276 AU), and 1.76$^{\prime\prime}$ (15136 AU) respectively. During the observation, J1733--1304, J1832--2039, and J1924--2914 were used as flux calibrator, phase calibrator, and bandpass calibrator respectively. The systemic velocity ($V_{LSR}$) of G10 is $\sim$68.50 km s$^{-1}$ \citep{rol11}.

For initial data reduction and spectral imaging, we used the Common Astronomy Software Application ({\tt CASA 5.4.1}) with the standard ALMA data reduction pipeline \citep{m07}. During the analysis of the raw data of G10, the continuum flux density of the flux calibrator for each baseline was scalled and matched with the Perley-Butler 2017 flux calibrator model with 5\% accuracy using task {\tt SETJY} \citep{pal17}. We made flux and bandpass calibration after discarding the bad data using {CASA} pipeline with task {\tt hifa\_bandpassflag} and {\tt hifa\_flagdata}. After the initial calibration, we split the target data set into another data set using task {\tt MSTRANSFORM} with all available rest frequencies. For the continuum subtraction procedure, we used the task {\tt UVCONTSUB} in the UV plane of split calibrated data. We generated the spectral datacubes of G10 using task {\tt TCLEAN} for each rest frequency with a Briggs weighting robust parameter of 0.5. For the correction of the primary beam pattern in the synthesized image, we applied the {\tt IMPBCOR} task in CASA. Earlier, \cite{gor20} presented a detailed description of the observations of the hot molecular core G10 using ALMA.

\begin{figure*}
	\centering
	\includegraphics[width=0.98\textwidth]{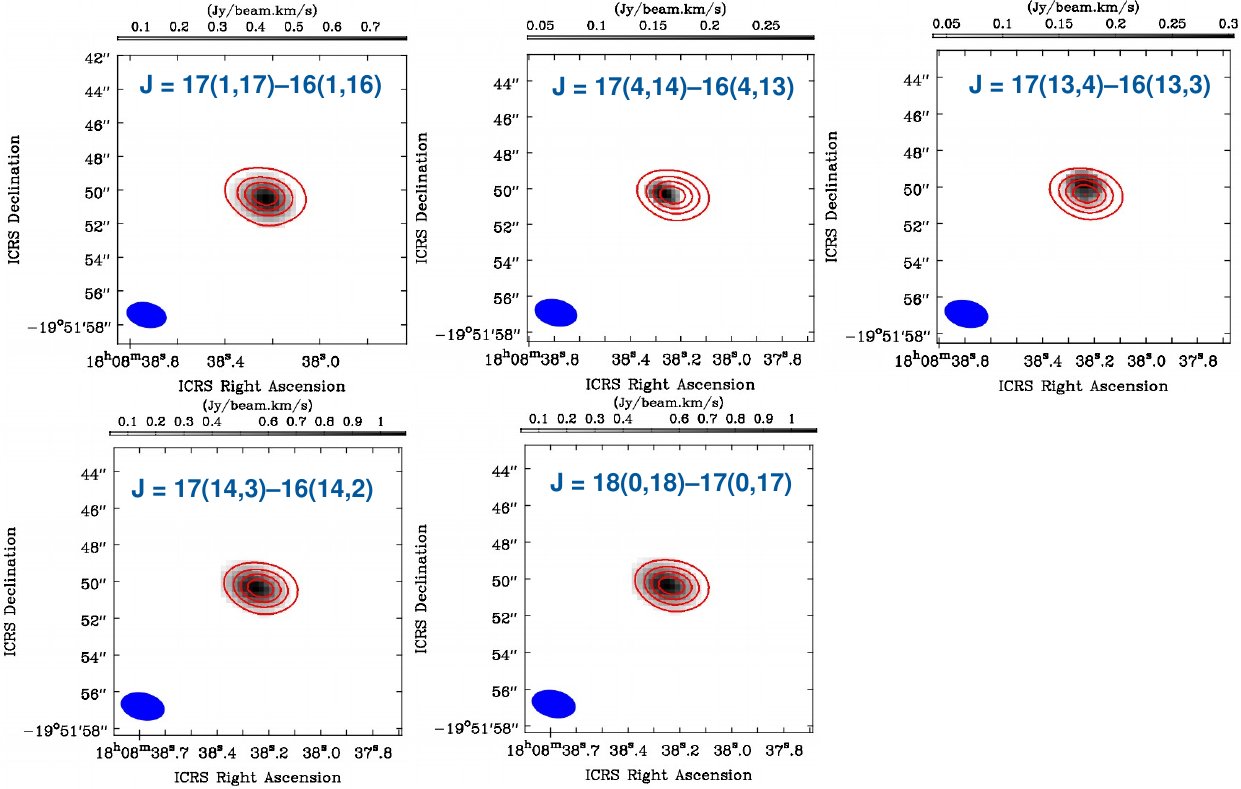}
	\caption{The integrated emission map of unblended transitions of AAN towards G10 which was overlaid with the 1.87 mm continuum emission (red contour). Contour levels are at 20\%, 40\%, 60\%, and 80\% of the peak flux. The blue circle indicated the synthesized beam of the integrated map.}
	\label{fig:emi}
\end{figure*}

\section{Result}
\label{ref}
\subsection{Analysis of the emission lines of interstellar AAN towards G10}
 We extracted the sub-millimeter spectra of hot molecular core G10 from the continuum subtracted spectral data cubes to make a 2.5$^{\prime\prime}$ diameter circular region at the center of RA (J2000) = (18$^{h}$08$^{m}$38$^{s}$.232), Dec (J2000) =  (--19$^\circ$51$^{\prime}$50$^{\prime\prime}$.440). The systematic velocity of the sub-millimeter spectra towards G10 was 68.50 km s$^{-1}$. After the extraction of the sub-millimeter spectrum, we used CASSIS \citep{vas15} for the identification of the emission lines of interstellar molecules towards G10. After careful spectral analysis, we identified the emission lines of the complex organic molecule AAN (a-type), which was known as a possible precursor of glycine. We used  the Cologne Database for Molecular Spectroscopy (CDMS)\footnote{\href{https://cdms.astro.uni-koeln.de/cgi-bin/cdmssearch}{https://cdms.astro.uni-koeln.de/cgi-bin/cdmssearch}} \citep{mu05} to identify the emission lines of AAN. The hot molecular core G10 was chemically very rich and the detection of AAN was extremely difficult due to contamination of other nearby molecular lines. We detected a total of twenty-one rotational transition lines of AAN between the frequency ranges of 147.55--148.01 GHz, 148.50--149.43 GHz, 153.00--153.93 GHz, 154.0--154.93 GHz, and 158.49--159.43 GHz towards G10.

 After the identification of emission lines of AAN from the submillimeter spectra of G10, we fitted the Gaussian model over the observed spectra of AAN using the line analysis module in CASSIS. After the fitting of Gaussian model over the observed spectra of AAN, we estimated the Full-Width Half Maximum (FWHM), quantum numbers ({${\rm J^{'}_{K_a^{'}K_c^{'}}}$--${\rm J^{''}_{K_a^{''}K_c^{''}}}$}), upper state energy ($E_u$), Einstein coefficients ($A_{ij}$), peak intensity and integrated intensity ($\rm{\int T_{mb}dV}$). There was no missing a-type transition of AAN in this data. The summary of the observed transitions of AAN and Gaussian fitting of spectral parameters were presented in Table~\ref{tab:MOLECULAR DATA} and the observed spectra of AAN with the best fitting Gaussian model were shown in Figure~\ref{fig:emission}. Earlier, the emission lines of AAN were first detected from Sgr B2(N) using the IRAM 30 m single-dish radio telescope \citep{bel08}. The spectral linewidth of observed emission lines of AAN towards Sgr B2(N) was $\sim$7 km s$^{-1}$ \citep{bel08}. So, there was a high probability of the contamination of other nearby molecular transitions with AAN towards Sgr B2(N) whereas our detected spectral width of AAN towards G10 was found between the range of 4--6 km s$^{-1}$. Our observed maximum transition lines of AAN towards G10 were blended with other nearby molecular lines because the observed transitions of AAN did not resolve due to the low spectral resolution. Our identified transitions of AAN between the frequency range of 147.55--148.01 GHz and 154.0--154.93 GHz towards G10 were also detected towards Sgr B2(N) by IRAM \citep{bel08}.

\begin{table*}{}
	\centering
	\caption{Summary of 2D Gaussian fitting parameters over the integrated emission map of AAN 
	}
	\begin{adjustbox}{width=0.8\textwidth}
		\begin{tabular}{cccccccccccc}
			\hline
			Transition& Integrated flux&Peak flux&Emitting region&Position angle\\
			
			[${\rm J^{'}_{K_a^{'}K_c^{'}}}$--${\rm J^{''}_{K_a^{''}K_c^{''}}}$]& [Jy.km s$^{-1}$]    &[Jy beam$^{-1}$. km s$^{-1}$]&[$^{\prime\prime}$]&[$^{\circ}$]\\
			\hline
			17(1,17)--16(1,16)&1.179$\pm$0.29&0.826$\pm$0.02&1.12&77.960\\
			17(4,14)--16(4,13)&0.830$\pm$0.06&0.733$\pm$0.04&1.13&77.971\\
			17(13,4)--16(13,3)&0.741$\pm$0.09&0.607$\pm$0.03&1.11&77.960\\
			17(14,3)--16(14,2)&1.805$\pm$0.59&1.248$\pm$0.65&1.12&77.968\\
			18(0,18)--17(0,17)&1.386$\pm$0.76&1.174$\pm$0.98&1.12&77.968\\
			
			\hline
		\end{tabular}	
	\end{adjustbox}
	
	
	\label{tab:prop}
\end{table*}

\subsection{Spatial distribution of AAN}
 After the identification of the emission lines of AAN towards G10, we extracted the integrated emission map of AAN using task {\tt IMMOMENTS} in CASA, which was shown in Figure~\ref{fig:emi}. The integrated emission map of AAN was overlaid with the 1.87 mm continuum emission map of G10\footnote{\cite{gor20} discussed the continuum emission towards G10 using ALMA}. We also observed that the emission map of AAN has a peak at the position of the continuum. The integrated emission maps were generated by integrating the spectral data cubes in the velocity range where the emission line of AAN was detected. We generated the integrated emission map only for the unblended transition of AAN towards G10. The integrated emission map indicated that the AAN molecule arises from the warm inner region of the hot core. We estimated the emitting region of AAN by fitting the 2D Gaussian over the integrated emission map of AAN using the CASA task {\tt IMFIT} towards the G10.
  The deconvolved beam size of the emitting region was calculated by the following equation\\

$\theta_{S}=\sqrt{\theta^2_{50}-\theta^2_{\rm beam}}$~~~~~~~~~~~~~~Eq.~1\\

where $\theta^2_{50} = 2\sqrt{A/\pi}$ was the diameter of the circle whose area ($A$) was enclosing $50\%$ line peak and $\theta_{beam}$ was the half-power width of the synthesized beam \citep{riv17, mondal21}. The estimated emitting region of AAN with transitions J = 17(1,17)--16(1,16), J = 17(4,14)--16(4,13), J= 17(13,4)--16(13,3), J = 17(14,3)--16(14,2), and J = 18(0,18)--17(0,17) were 1.12$^{\prime\prime}$, 1.13$^{\prime\prime}$, 1.11$^{\prime\prime}$, 1.12$^{\prime\prime}$, and 1.12$^{\prime\prime}$ respectively. So, the emitting region of AAN towards G10 varied between 1.11$^{\prime\prime}$--1.13$^{\prime\prime}$. After the fitting of the 2D Gaussian over the integrated emission map of AAN, we also estimated the integrated flux, peak flux, and position angle, which are presented in Table~\ref{tab:prop}. We noticed that the emitting region of AAN is smaller than the synthesized beam size, which indicates the unblended transition lines of AAN are not well spatially resolved or, at best, marginally resolved.

 \subsection{Rotational diagram analysis of AAN}
\label{rotd}
 In this work, we have detected the multiple hyperfine transition lines of AAN towards G10. The rotational diagram method is one of the best ways to obtain the column density ($N$) in cm$^{-2}$ and rotational temperature ($T_{rot}$) in K of the detected emission lines of AAN. We assumed that the observed AAN spectra were optically thin and that they obeyed the Local Thermodynamic Equilibrium (LTE) conditions. The assumption of the LTE condition was reasonable towards the G10 because the density of the inner regions of the hot core was $\sim$7$\times$10$^{7}$ cm$^{-3}$ \citep{rol11}. The equation of column density can be written as in the case of optically thin lines \citep{gold99},\\\\

${{N_u^{\rm thin}}=\frac{3{g_u}k_B\int{T_{mb}dV}}{8\pi^{3}\nu S\mu^{2}}}$~~~~~~~~~~~~~~Eq.~2\\\\
 where, $k_B$ is the Boltzmann constant, $\rm{\int T_{mb}dV}$ is the integrated intensity, $\mu$ is the electric dipole moment, $g_u$ is the degeneracy of the upper state, $\nu$ is the rest frequency, and the strength of the transition lines were indicated by $S$. The total column density of detected species under LTE conditions can be written as,\\\\

$\frac{N_u^{\rm thin}}{g_u} = \frac{N_{\rm total}}{Z(T_{\rm rot})}\exp(-E_u/k_BT_{\rm rot})$~~~~~~~~~~~~~~Eq.~3\\\\
 where ${Z(T_{rot})}$ is the partition function at extracted rotational temperature ($T_{\rm rot}$). The rotational partition function at 75 K is 4403 and 150 K is 12460, respectively. The upper state energy of the observed molecular lines defined as $E_u$. In another way, the Eq.~3 can be rearranged as,\\\\

\begin{figure}
	\centering
	\includegraphics[width=0.5\textwidth]{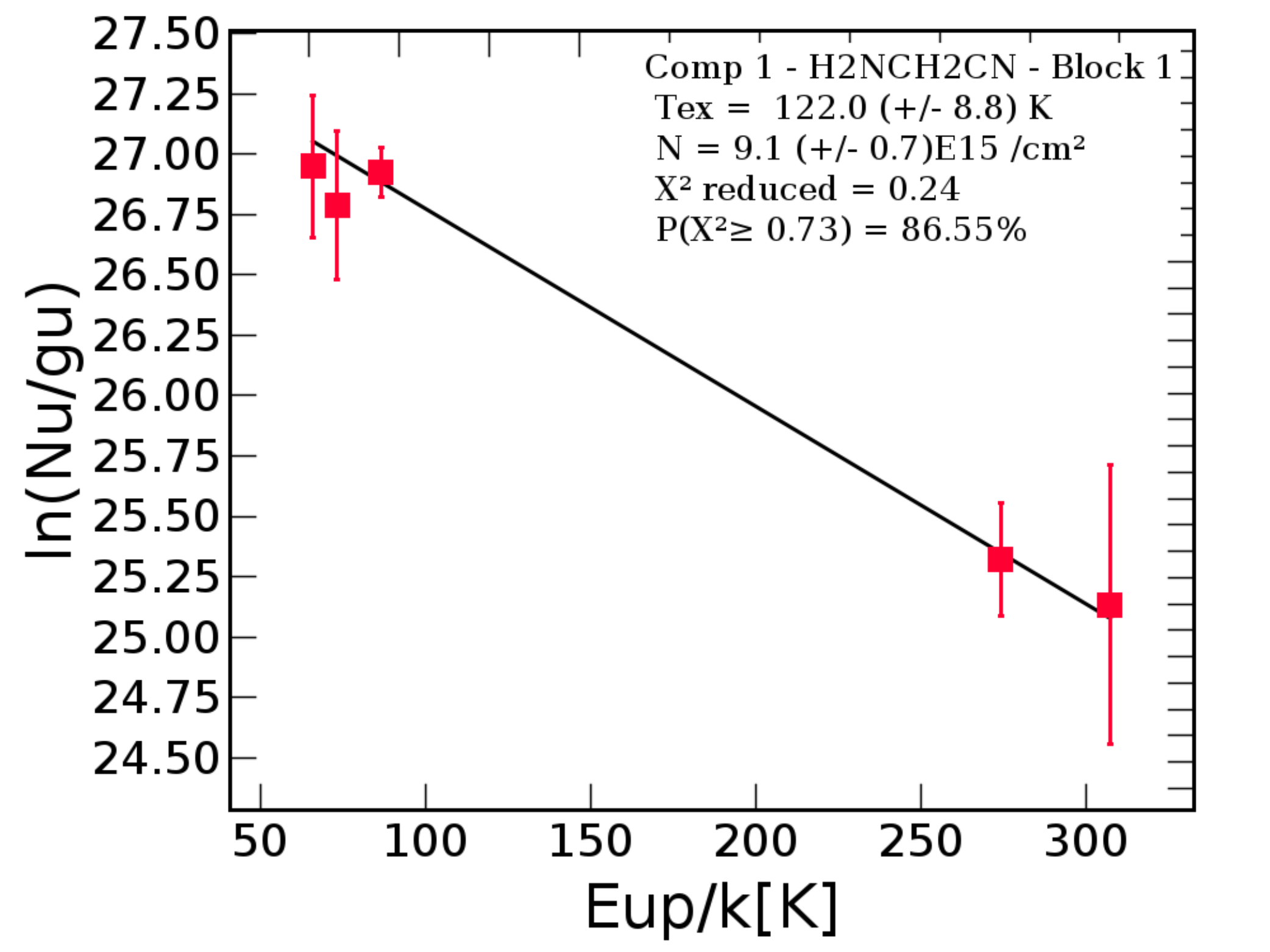}
	\caption{The rotational diagram of AAN towards G10. The red filled squares indicate the optically thin approximation data points and the red lines represent the error bars. The best-fit column density and rotational temperature are mentioned in the image.}
	\label{fig:rotd} 
\end{figure}

$ln\left(\frac{N_u^{\rm thin}}{g_u}\right) = ln(N)-ln(Z)-\left(\frac{E_u}{k_BT_{rot}}\right)$~~~~~~~~~~~~~~Eq.~4\\\\
 Eq.~4 presented a linear relationship between the upper state energy (E$_{u}$) and $\ln(N_{u}/g_{u}$) of the observed complex organic molecule AAN. The $\ln(N_{u}/g_{u}$) were estimated from the Eq.~2. From Eq.~4, it is evident that spectral data points with respect to different transition lines of AAN should be fitted with a straight line whose slope is inversely proportional to $T_{rot}$, with its intercept yielding $\ln(N/Z)$, which in turn will help to estimate the molecular column density. During the rotational diagram analysis, we extracted the line parameters such as FWHM, upper energy ($E_u$), Einstein coefficients (A$_{ij}$), and integrated intensity ($\int T_{mb}dV$) using a Gaussian fitting over the observed transitions of AAN using the CASSIS. The fitting parameters are presented in Table~\ref{tab:MOLECULAR DATA}. We observed many transitions of AAN are blended as the present spectral resolution is not sufficient to resolve the hyperfine components. We used the most intense hyperfine transitions of AAN in our rotational diagram analysis. We have observed the maximum transitions of AAN blended with nearby molecular transitions. So, we used only unblended transitions of AAN during the rotational diagram analysis. The computed rotational diagram was shown in Figure~\ref{fig:rotd}. In the rotational diagram, the red error bars represent the absolute uncertainty of $\ln(N_{u}/g_{u}$) originating from the error of the estimated integrated intensity, which was calculated after fitting the Gaussian model over the detected unblended emission lines of AAN. After the rotational diagram analysis, the estimated column density of AAN was (9.10$\pm$0.7)$\times$10$^{15}$ cm$^{-2}$ with rotational temperature ($T_{rot}$) 122$\pm$8.8 K. The estimated fractional abundance of AAN was 7.01$\times$10$^{-8}$ where column density of \ce{H2} towards G10 was 1.3$\times$10$^{23}$ cm$^{-2}$ \citep{suz16}.

\section{Discussion}
\label{dis}

\subsection{AAN in the G10}
  The emission lines of AAN towards the hot molecular core G10 were the first time detected using ALMA with an estimated source size of 1.11$^{\prime\prime}$--1.13$^{\prime\prime}$ which is presented in this article. After the spectral analysis using the Gaussian model, it was found that the maximum transition lines of AAN were blended with nearby molecular lines as presented in Table~\ref{tab:MOLECULAR DATA}. We observed that the J = 17(1,17)--16(1,16), 17(4,14)--16(4,13), 17(13,4)--16(13,3), 17(14,3)--16(14,2), and 18(0,18)--17(0,17) transition lines of AAN were not blended with nearby other molecular lines, and these transition lines were used during rotational diagram analysis to estimate the column density and gas temperature of AAN. The line contamination of AAN with other molecular lines towards G10 was less with respect to Sgr B2(N) because the estimated source size of G10 was 1.11$^{\prime\prime}$--1.13$^{\prime\prime}$ whereas the source size of Sgr B2(N) was 2$^{\prime\prime}$. The estimated column density of AAN towards G10 was (9.10$\pm$0.7)$\times$10$^{15}$ cm$^{-2}$ with rotational temperature ($T_{rot}$) 122$\pm$8.8 K, and corresponding line width 4--6 km s$^{-1}$ with a centroid velocity 68.50 km s$^{-1}$. The fractional abundance of AAN with respect to \ce{H2} was 7.01$\times$10$^{-8}$ where the column density of \ce{H2} was 1.3$\times$10$^{23}$ cm$^{-2}$ \citep{suz16}. The estimated rotational temperature indicated that the AAN molecule was mainly coming from the hot core of G10 because the temperature of the hot core was above 100 K \citep{rol11}. Earlier, the emission lines of AAN were detected from another hot molecular core Sgr B2(N) with fractional abundance 2.2$\times$10$^{-9}$ \citep{bel08}. The fractional abundance of AAN in G10 was $\sim$10 times higher than Sgr B2(N). In Sgr B2(N), the emission lines of AAN arise from the hot core region, which is also known as the `large molecule heimat' \citep{bel08}. Recently, \citet{mel20} detected the emission lines of AAN from Sgr B2(N1S) using ALMA with ReMoCA spectral line survey and they estimated the column density of AAN was 1.1$\times$10$^{17}$ cm$^{-2}$ with excitation temperature 200 K. The estimated column density of AAN derived from LTE modelling of ReMoCA spectral line survey toward Sgr B2(N1S) was 1.1$\times$10$^{17}$ cm$^{-2}$, which was $\sim$3.9 times higher than the column density reported in \cite{bel08} for Sgr B2(N) from observations with the IRAM 30 m telescope. During the analysis of the column density of AAN, \citet{bel08} and \citet{mel20} both assumed the same emission size. There are several reasons for this difference. First of all, \citet{bel08} did not account for the contribution of the vibrational partition function. They assumed the temperature of AAN was 100 K and this contribution amounts to a factor of 1.09 and after accounting for this, the column density derived from the IRAM 30 m data becomes 2.8$\times$10$^{16}$ cm$^{-2}$.

\begin{table*}
	\caption{Comparison between simulated and observed fractional abundances of AAN}\label{tab:comparison} 
	\centering      
	\begin{tabular}{c|c|c|c|cc}
		\hline 
		& \multicolumn{3}{c}{Simulated Values$^{\rm a}$} & \multicolumn{2}{c}{Observed Values$^{\rm b}$} \\
		\hline
		Species & Fast & Medium & Slow & G10\\
		\hline
		&$f$(AAN)~~~~~~~T (K)&$f$(AAN)~~~~~~~T (K)&$f$(AAN)~~~~~~T (K)& $f$(AAN)~~~~~~~~T (K)\\
		\hline
		\ce{NH2CH2CN} & $4.3\times10^{-9}$~~123 & $1.2\times10^{-8}$~~~122 & $4.7\times10^{-9}$~~~118 & $7.01\times10^{-8}$~~~122 \\
		\hline
	\end{tabular}
	
	
	Notes: a -- Values taken from Table\,8 of \cite{gar13}; \\b --  this work.
	\label{tab:comp}
\end{table*}


\subsection{Comparison with observation and simulated abundances of AAN}

We compared our estimated abundance of AAN with the three-phase warm-up modelling results of \cite{gar13}. \cite{gar13} assumed that there would be an isothermal collapse phase after a static warm-up phase. Under the free-fall collapse, the density increases from $n_{H}$ = 3$\times$10$^{3}$ to 10$^{7}$ cm$^{-3}$ in the first phase, and the temperature of the dust reduces to 8 K from 16 K. The temperature fluctuates from 8 K to 400 K in the second phase, but the density remains constant at $\sim$10$^{7}$ cm$^{-3}$ \citep{gar13}. The temperature of G10 was $\sim$150 K, which is a typical hot core temperature, and the number density ($n_{H}$) of this source is $\sim$10$^{7}$ cm$^{-3}$ \citep{rol11, gor20, ohi19, suz16}. Thus, the hot core model of \cite{gar13} is suitable for understanding the chemical evolution of AAN towards the G10. \cite{gar13} used the fast, medium, and slow warm-up models based on the time scale. In Table~\ref{tab:comp}, we compared the observed fractional abundance of AAN with the three-phase warm-up model of \cite{gar13} and we observed that the simulation result was nearly similar to our observational results. We found that the medium warm-up model best matched the observation values of AAN towards the G10. During the comparison of the simulated and observed abundances of AAN, we found a tolerance of factor 7, which is a reasonably good agreement. Earlier, \citet{ohi19} and \citet{suz16} claimed that the \ce{CH2NH} molecule obeys the medium warm-up model towards G10, and we observed that the AAN molecule also obeys the medium warm-up model towards G10. So, we conclude that both \ce{CH2NH} and AAN molecules obey the medium warm-up model towards G10. It is indicated that the \ce{CH2NH} molecule plays as a possible precursor of AAN towards G10. Earlier, another glycine precursor molecule, methylamine \ce{(CH3NH2)} was also detected with column density (1.0$\pm$0.7)$\times$10$^{15}$ cm$^{-2}$ from the G10 using the Nobeyama 45 m radio telescope. The detection of three possible glycine precursors like AAN (our work), \ce{CH3NH2} \citep{ohi19}, and \ce{CH2NH} \citep{suz16} towards G10 gives confidence in the possibility of the presence of interstellar glycine in this hot molecular core.

\cite{gar13} proposed that AAN is produced mainly by the addition of radical \ce{CH2CN} with \ce{NH2}, which is formed by hydrogen abstraction from \ce{CH3CN} molecules by \ce{NH2} at around 60--80 K temperature, resulting in the fractional abundance of around 10$^{-9}$ in the grain surface. Further, the evaporation of \ce{CH3CN} from the grains at a temperature of nearly 90 K produces \ce{CH2CN} with an efficiency of 50\% with gas-phase protonation and electron recombination reactions. When this radical re-accretes to the grain surface and reacts with \ce{NH2} then it produces a fractional abundance of AAN in the order of 10$^{-8}$. The estimated fractional abundance of AAN towards the G10 was 7.01$\times$10$^{-8}$ which indicated that the proposed chemical modeling of \cite{gar13} satisfied the environment of G10.

\subsection{Searching of interstellar glycine towards G10} 
 After the successful detection of AAN, we also looked for the emission lines of glycine towards G10. After the deep searches of glycine conformer I and II using the LTE module in CASSIS, we did not detect this molecule within the limits of our LTE analysis. Using the same parameter of AAN ($\theta_{S}\sim$1.12$^{\prime\prime}$ and $T_{ex}$ = 122 K), the estimated upper limit column density of glycine conformer I was $\leq$1.25$\times$10$^{15}$ cm$^{-2}$ and conformer II was $\leq$4.86$\times$10$^{13}$ cm$^{-2}$. The energy of glycine conformer I is 705 cm$^{-1}$ (1012 K) lower than that of glycine conformer II \citep{lov95}. The dipole moment of glycine conformer I is $\mu_{a}$ = 0.911 D and $\mu_{b}$ = 0.607 D, whereas conformer II is $\mu_{a}$ = 5.372 D and $\mu_{b}$ = 0.93 D \citep{lov95}. In the interstellar medium, the a-type transitions of glycine are more prominent because the molecular line intensity is proportional to the square of the dipole moment \citep{lov95}.

After the detection of \ce{CH3NH2} and \ce{CH2NH} towards G10, \citet{ohi19} claimed that the hot molecular core G10 was one of the sources in an interstellar medium where glycine would be detectable. Our detection of AAN in G10 using ALMA gives more confidence that the hot molecular core G10 has the ability to form glycine. Earlier, \cite{ohi19} estimated the upper limit column density of glycine conformer I with respect to \ce{CH3NH2} was $\leq$1.1$\times$10$^{15}$ cm$^{-2}$ which satisfied our estimated upper limit column density of glycine with respect to AAN. We propose to conduct a survey of glycine conformer I and II and its precursors (\ce{CH3NH2}, \ce{CH2NH}, AAN, and \ce{NH2CH2C(O)NH2}) molecules towards the hot molecular core G10 with a higher integration time and better spectral resolution to solve the puzzle of the glycine lines in the interstellar medium.
\section{Conclusion}
\label{conclu}
 In this article, we presented the detection of the glycine nitrile molecule AAN towards G10 using ALMA band 4. The main results are as follows.\\\\
 1. We successfully detected a total of twenty-one rotational emission lines of AAN towards the hot molecular core G10 using ALMA band 4 observation.\\\\
 2. The derived column density of AAN using the rotational diagram method was (9.10$\pm$0.7)$\times$10$^{15}$ cm$^{-2}$ with rotational temperature 122$\pm$8.8 K. The relative abundance of AAN with respect to \ce{H2} was 7.01$\times$10$^{-8}$ where column density of \ce{H2} in hot molecular core G10 was 1.3$\times$10$^{23}$ cm$^{-2}$ \citep{suz16}.\\\\
 3. In the discussion, we compared our fractional abundance of AAN with the three-phase warm-up model by \cite{gar13} and we found that the medium warm-up model satisfied our AAN abundance. \\\\
 4. After the detection of AAN in G10, we also looked for the emission lines of glycine conformers I and II. Within the limits of our LTE analysis, we did not detect the emission lines of glycine conformer I and II towards G10. We calculated the upper limit column density of glycine conformer I as $\leq$1.25$\times$10$^{15}$ cm$^{-2}$ and conformer II as $\leq$4.86$\times$10$^{13}$ cm$^{-2}$. \\\\
 5. After the unsuccessful detection of glycine conformers I and II using ALMA, we conclude that the emission line of glycine may be below the confusion limit in G10.
	
	\section*{Acknowledgement}
We thanks both anonymous referee for helpful comments that improved the manuscript. The plots within this paper and other findings of this study are available from the corresponding author upon reasonable request. This paper makes use of the following ALMA data: ADS /JAO.ALMA\\\#2017.1.00121.S. ALMA is a partnership of ESO (representing its member states), NSF (USA), and NINS (Japan), together with NRC (Canada), MOST and ASIAA (Taiwan), and KASI (Republic of Korea), in cooperation with the Republic of Chile. The Joint ALMA Observatory is operated by ESO, AUI/NRAO, and NAOJ.
	
	\section*{Funding}
	This research did not receive any specific grant from funding agencies in the public, commercial, or not-for-profit sectors.
	\section*{Declaration of Competing Interest}
	The authors declare that they have no known competing financial interests or personal relationships that could have appeared to influence the work reported in this paper

\end{document}